\documentstyle[12pt]{article}

\newcommand{\eq}[1]{Eq.(\ref{#1})}
\def\d{\partial}
\def\a{\alpha}
\def\ll{(a\rho)^{-1/2}}
\def\be{\begin{equation}}
\def\ee{\end{equation}}

\begin{document}
\vspace{0.5in}
\begin{center}

{\Large Two-dimensional Bose gas at low density } \\
\vspace{0.4in}

{\large A.A.~Ovchinnikov} \\
\vspace{0.2in}
{\it Institute for Nuclear Research of the Russian Academy
of Sciences,\\ Moscow 117312 Russia} \\
\vspace{0.7in}

\end{center}
\begin{abstract}

We propose a new method to describe the interacting bose gas at zero
temperature.
For three-dimensional system the correction
to the ground- state energy in density is reproduced.
For two-dimensional dilute
bose gas the ground- state energy in the leading order in the
parameter $|\ln\a^2\rho|^{-1}$ where $\a$ is a
scattering length is obtained.

\end{abstract}
\newpage

At present time two-dimensional models attract much attention
in connection with the problems of high-$T_c$ superconductivity and
fractional quantum Hall effect.
In this context the description of the two-dimensional system of
bosons may be important.
For instance the system of planar fermions in the magnetic
field is equivalent to the system of the interacting bosons with the
additional long-range Chern-Simons interaction. The problem of the
hard core bosons on a lattice at high density is closely related to
the description of different strongly correlated electronic systems.
The conventional methods of the description of the
interacting bose gas are inapplicable in both of the above mentioned
problems.
In this context the new ways to treat the system in the two different
physical limits where the perturbation theory is possible are of
interest.

Various methods to describe the system of bosons with the
pairwise interaction were
introduced which in one way or another consisted in summing up an
infinite subset of the terms in the perturbation series.
The example of this procedure is the Bogoliubov's method \cite{B}
which is correct in the high density limit which means that
the range of the potential is much larger than the average
particle spacing.
In the opposite limit of the dilute bose gas the perturbation theory
\cite{B} does not work which manifests in the divergence in the
expression for the energy for the $\delta$-function type potential.
Various modifications of this procedure using the pseudopotential or
the diagrammatic expansion methods, for example, were proposed
\cite{LY}.  However the reason why the procedure of ref.\cite{B} can
be applied when the two-body potential is not small and the wave
function is not close to the unperturbed one is obscured.  The
generalization of this methods to the 2D system is not straightforward
due to the behavior of the modified scattering amplitude which
approaches zero in the low energy limit.  In fact the methods
\cite{LY} can be considered as an application of the Bogoliubov's
approach to a system with the parameters chosen in such a way that
both this approximation is valid and at the same time the density
$\rho$ is small in a sense $\a\rho^{1/d}<<1$ ($\a$ is the scattering
length).  Although this can be a basis for the
solution of 2D problem the estimate of the accuracy of the
approximation \cite{B} for a given potential is required.

In the present paper we suggest a new method to describe the
system of bosons for the two limiting cases and investigate the
ground state properties of 2D bose gas at zero temperature. Our
method is closely related to the  approach proposed by Lieb \cite{L}
which makes use of the distribution functions related to the ground
state wave function.
We use the decomposition of the logarithm of the wave function into a
series over the $n$-particle functions.
We argue that at small density the proposed
expansion corresponds to the expansion of energy in the small
parameter. Although in practice
it is difficult to solve the corresponding system of equations beyond
the leading order approximation which in the framework of our approach
corresponds to the wave function
of the Jastrow form, the convincing arguments in favor of the
validity of this procedure in the leading order can be found.
Of course the distribution functions can be calculated using the
wave function of the Jastrow form.
In this way the connection with the approach of ref.\cite{L} can be
made.

The equation for the energy $E_0$ and the wave function
$\Phi(x_1,\ldots x_N)$ of the ground state for a system of $N$
particles in the volume $V$ interacting with the two-body potential
$U(x)$ has the form
\be
\left(-\sum_{i}\d_i^2+2\sum_{i<j}U(x_{ij})\right)\Phi=E_0 \Phi,
\hspace{0.2in}
i,j=1,\ldots N,
\label{1.0}
\ee
where $\d_{i}=\d/\d x_{i}$, $x_{ij}=x_{i}-x_{j}$
and we denote by $x_i$ the $d$- dimensional space vector of
$i$-th particle throughout the paper.
The ground state wave function $\Phi(x_1,...x_N)$ is a
symmetric and positive function of its arguments. According to
\cite{Bijl} (see also \cite{BZ}) one can seek for the function $\Phi$
in the form
\[
\Phi(x_{1},\ldots x_{N})=\exp S(x_{1},\ldots x_{N}).
\]
We observe that the function $S(x_{1},\ldots x_{N})$ can be
expanded as follows
\be
S(x_{1},\ldots x_{N})=\sum_{i<j}S_{2}(x_{ij})+
                      \sum_{i<k<l}S_{3}(x_i,x_k,x_l) + \ldots ,
\label{1.1}
\ee
where
$S_{n}(x_1,\ldots x_n)$
are the symmetric functions depending on the relative interparticle
distances and subjected to the constraints
\be
\int dx_{n} S_{n}(x_1,\ldots x_n) = 0, \hspace{0.2in} n>2.
\label{1.2}
\ee
Integrating \eq{1.1} over the coordinates of $N-n$ particles
subsequently for $n=2,3\ldots$ and using the condition (\ref{1.2})
one can show that the expansion (\ref{1.1}) is an irreducible one
which means that for a given function
$S(x_{1},\ldots x_N)$ the set of the functions $S_n$ is unique.
Substituting \eq{1.1} into the equation (\ref{1.0}) we
obtain the equation
\be
E_0 = \sum_{i\neq j} F(x_{ij}) - \sum_{i\neq k\neq l}
\d S(x_{ik}) \d S(x_{il}) + \ldots,
\label{1.3}
\ee
\[
F(x)= -\d^2 S(x) - \d S(x) \d S(x) + U(x),
\]
where the terms depending on the pair function $S_{2}(x)=S(x)$ only
are indicated explicitly. Integrating \eq{1.3} over the coordinates
$n+1,\ldots N$ and using \eq{1.2} we obtain the $n$-particle equation
for the functions $S_n$.
In general the system of the equations cannot be solved and the
assumptions on the higher order $S_{n}$ - functions are required.
For the dilute bose gas the perturbation theory
in the small parameter can be used.
Although the calculations in the high orders are too complicated
one can argue that the expansion of the energy in the small parameter
corresponds just to the expansion (\ref{1.1}).
Here we will demonstrate it in the lowest order
which corresponds to the pairwise wave function.
It is plausible that the same is valid for the higher order terms.
Suppose for a while that only the function $S(x)$ is not equal to
zero. Integrating the equation (\ref{1.3}) over the coordinates
$1,\ldots N$ we obtain for the parameter $a$ defined as
\[
E_0 = N\rho a
\]
($\rho=N/V$, $E_0/V=a\rho^2$) the equation
\be
     a = \frac{N-1}{N} \int dx F(x)
\label{1.4}
\ee
which is equivalent to
$a=\int dx (U-(\d S)^2)$ in the infinite volume limit. Integrating
\eq{1.3} over the coordinates $3,\ldots N$ and using \eq{1.4} we
obtain the equation
\be
\d^{2}S(x) + \d S \d S(x) - U(x) -
\rho \d_{1}^{2} \int dx_3 S(x_{13}) S(x_{23}) = - a/V.
\label{1}
\ee
where the notation $x=x_{12}$ is used and the terms of order
$\sim 1/V$ are retained. The right- hand side is equal
to zero in the limit $V\rightarrow \infty$.
We assume the periodic boundary conditions so that the integral over
the total derivative is zero.  Due to the condition (\ref{1.2}) the
contribution of $S_3$ to \eq{1.4} is of the form
$\rho\int dx_{1}dx_{2}(\d_{1}S(123))^2$
and the example of its contribution to \eq{1} is
$\rho\int dx_{3}\d_{3}S(123)\d_{3}S(13)$.
One can use the solution of \eq{1} to estimate the function $S_3$ with
the help of the three-particle equation and show that these integrals
are suppressed.  We will show below that it is the smallness of $S(x)$
at the distances $\sim\rho^{-1/d}$ ($d=2,3$) that allow one to neglect
the $S_3$ function in the lowest order approximation.
The equation (\ref{1}) reproduce the correction in density
\cite{LY} to the ground- state energy for 3D system and predicts the
leading order term for 2D system.

Note also that \eq{1} can be regarded as an equation for the
trial variational wave function of the Jastrow form. In this case it
follows from the form of the solution of \eq{1} that the variational
energy is given by \eq{1.4} with the accuracy up to the higher order
terms in the expansion parameter.

If the second term in the left-hand side of the equation (\ref{1}) can
be neglected then the equation can be easily solved using the Fourier
transformation:
\[
\rho S_k^2-k^2 S_k-U_k=0, \hspace{0.3in} S_k=\int dx e^{ikx} S(x).
\]
Substituting the solution of this equation $S_k$ to the
expression for the energy
\[
  a=U_0 + \int dx (\d S(x))^2 = U_0 + \int_k k^2 S_k^2 ,
\]
($U_0=\int dxU(x)$) we get
\be
a=U_0+\frac{1}{2\rho^2}\int_k
\left( \left(k^4+4U_{k}\rho k^2\right)^{1/2}-k^2-2U_{k}\rho \right),
\label{Bog}
\ee
where the notation $\int_k=\int d^{d}k/(2\pi)^d$ is used.
\eq{Bog} is the Bogoliubov's expression for the energy. One can
estimate the accuracy of \eq{Bog} for a given two-body potential
$U(x)$. The corrections are determined by the function $S(x)$ which
should be small in order the approximation (\ref{Bog}) to be
valid.
For instance in 2D for the potential which is a repulsive square well
in momentum space, $U_k=u,~k<b^{-1}$, $U_k=0,~k>b^{-1}$, the energy
is: ($i$) at $u\rho b^2>>1$, $a=u(1-{1\over 4\pi}{1\over \rho b^2})$,
$S(0)=-(u/\rho b^2)^{1/2}/2\pi$; ($ii$) at $u\rho b^2<<1$,
$a=u(1-{u\over 4\pi}\ln (u\rho b^2e^{1/2}))$,
$S(0)=-1/8\pi\rho b^2$.
Briefly, the potential should be relatively shallow compared to its
width and the spatial range should be much larger than the average
particle spacing.

Let us proceed with the solution of the equation (\ref{1})
in the low density limit. The particle density is assumed to be small
so that the dimensionless parameter $\a\rho^{1/d}<<1$. In this formula
$\a$ is the scattering length for the potential $U$ and $\rho^{-1/d}$
is the average interparticle distance in $d$- dimensional space.
The expansion parameters are respectively $|\ln\rho\a^2|^{-1}$
and $\a^{3/2}\rho^{1/2}$ in two and three dimensions.

\vspace{0.25in}

{\bf 1. ~~Three-dimensional system.}

Let us begin with the case of three spatial dimensions $d=3$.
Define $S_0(x)$ as the solution of the equation (\ref{1}) at $\rho=0$:
\be
\d^2 S_0 + \d S_0 \d S_0-U=-\a/V.
\label{S0}
\ee
Here $\a$ is the unknown parameter and the periodic boundary
conditions are implied.
In terms of the function $\phi_0(x)=\exp(S_0(x))$ keeping the volume
finite and taking into account the periodic boundary conditions
we find that
\[
\a=\int dx (U(x)+S_0\d^2 S_0(x)) = \int dx U(x)\phi_0(x)
\]
is proportional to the scattering length for the potential $U(x)$.
For the hard-sphere potential $\a$ is the radius of the potential.
In the infinite volume limit the equation has the form
$(-\d^2+U)\phi_0(x)=0$ which is to be supplemented by
the boundary condition
$\phi_0(r)\rightarrow 1$ at $r\rightarrow\infty$ ($r=|x|$).
At the distances much larger than the range of the potential the
solution is $\phi_0(r)=1-\a/r$.
For the energy we have
$a=\a+\int dx(S\d^2S-S_0\d^2S_0)$ or, equivalently,
\be
a=\a-\int_k (k^{2}S_k^2 - k^{2}S_{0k}^{2}).
\label{3da}
\ee
Make use of the Fourier transformation in the equations (\ref{1}) and
(\ref{S0}). In the region $r>>\a$ where the condition $S_0(x)<<1$
is satisfied we have $S_0(r)=-\a/r$ and the main contribution
to the integral $\int dx(\d S_0)^2$ comes from the
integration over the region of the small $r\sim\a$.
Up to the corrections in $\rho$ the same is true
for the analogous integral for $S(x)$.

Therefore the Fourier transform of the function $U-(\d S)^2$, and
the same function for $S_0$, are independent of the momentum $k$ at
$k<<1/\a$. At these values of $k$,~ $S_{0k}=-\a/k^2$ and $S_k$ is
given by the solution of the equation $\rho k^2S_k^2-k^2S_k-\a=0$.
Substituting these functions into \eq{3da} we obtain
\be
a=\a+\frac{1}{2\rho^2}\int_k \left( \sqrt{ k^4+4\a\rho k^2 }
-k^2-2\a\rho+\frac{2\a^2\rho^2}{k^2} \right)
\label{3int}
\ee
The integral in this expression converges at large $k$ and is
saturated at the values $k\sim(a\rho)^{1/2}<<1/\a$ which justifies our
assumption that $\int dxe^{ikx}(U-(\d S)^2)=\a$. The corrections to
this formula does not change the result in the approximation
considered. Evaluating the integral in \eq{3int} we obtain the well
known result \cite{LY} for the expansion of the ground state
energy in powers of density in 3D:
\be
a=\a\left(1+\frac{16}{15\pi^2} \a^{3/2} \rho^{1/2} \right).
\label{aa}
\ee

\vspace{0.25in}

{\bf 2. ~~Two-dimensional system.}

Let us proceed with the investigation of two-dimensional system.
In 2D it is easy to find the solution of \eq{1} and determine the
energy $a$ as the function of the density $\rho$ with the accuracy up
to the terms of the higher order in the small parameter
$a\sim|\ln\rho\a^2|^{-1}$.

As in three-dimensional case the solution of \eq{1} can be represented
in the form
\[
S_k =-\frac{1}{2\rho}\left(\sqrt{1+4\sigma_{k}\rho/k^2}-1\right),
\]
where $S_k$ is the Fourier transform of the function $S(x)$ and the
function $\sigma_k$ is defined by
\[
\sigma_{k}=\int dxe^{ikx}\left(U(x)-\d S \d S(x)\right).
\]
At the sufficiently small $k$ we have $\sigma_k=a$ while at the larger
$k$, $\sigma_k$ could depend on the momentum $k$. However it
can be shown that in fact with the accuracy of order $O(a^2)$
one can take $\sigma_k \simeq a$ for all $k<<\a^{-1}$. We are
interested in the behavior of the function $S(x)$ in the region
$\a<<r<<(a\rho)^{-1/2}$. As in 3D $\ll$ is the characteristic length,
the kind of correlation length in the problem. In this region $S(x)$
can be calculated as
\be
S(x)=-\frac{1}{2\rho} \int_k e^{ikx} \left(
\left(1+\frac{4a\rho}{k^2}\right)^{1/2} -1 \right)
\label{T2}
\ee
Evaluating this expression we find that
\be
S(x)=-a/4\pi+(a/2\pi)\ln\left((a\rho)^{1/2}r\right)
\label{T3}
\ee
which is valid in the indicated region with the accuracy up to the
terms of order of $a^2$. One can see from \eq{T2} that the function
$S(x)$ decreases quickly at the distances
much larger than the correlation length $\ll$:
\[
S(x)\simeq
 - \frac{1}{2\pi}\left(\frac{a}{\rho}\right)^{1/2}\frac{1}{r},
\hspace{0.25in} r>>\ll.
\]
Now using the equation (\ref{T3}) one can estimate the function
$\sigma_k$:
\[
|\sigma_k -a| < \frac{a^2}{2\pi}\ln\frac{k}{\sqrt{a\rho}}+O(a^2).
\]
This relation justifies
the initial assumption about the behavior of $\sigma_k$.
The other way to obtain the solution (\ref{T3}) is to represent the
equation (\ref{1}) in the form
\be
(\d^2-U(x))\phi(x)=\rho\phi(x)\d_1^2 \int dx_{3} S(x_{13})S(x_{23})-
(a/V)\phi(x)
\label{T4}
\ee
where $\phi(x)=\exp S(x)$.
The Fourier transform of the function $U(x)\phi(x)$ does not
depend on the momentum at $k<<1/\a$.
In the large distance region where
$S(x)<<1$ we can expand the function $\phi(x)\simeq 1+S(x)$ and
neglect the terms of order $S^{2}(x)$. Then the equation (\ref{T4})
has the solution of the same form as \eq{T3}.
The only difference is that the parameter $a$ is replaced by
the value of the integral $\int dxU(x)\phi(x)$.
To calculate it one has to integrate both sides of the equation
(\ref{T4}) over the space vector in the finite volume $V$:
\be
\int dx U(x)\phi(x) = a -
    \rho \int dx(1 - \phi(x)) \chi (x)
\label{T5}
\ee
\[
\chi (x_{12}) = \d_1^2 \int dx_3 S(x_{13})S(x_{23})
\]
In general the limit $V\rightarrow\infty$ of integral of some function
over the volume does not coincide with the integral of the limiting
function, because in the finite volume the function may have the
asymptotic $const./V$. This is indeed the case for $\int dx\chi(x)$
which is zero in the finite volume. However one can substitute the
product of the limiting functions  $(1-\phi(x))\chi(x)$ into \eq{T5}
since both functions approaches zero at $r\rightarrow\infty$.
Alternatively, one can use the equation (\ref{1}) directly in the
infinite volume limit.
The function $\chi(r)$ can be evaluated using \eq{T2}:
\[
\chi(r)\simeq aS(r),~(\a<<r<<\ll);
{}~~~\chi(r)\simeq
\frac{1}{2\pi}\left(\frac{a^{1/2}}{\rho^{3/2}}\right) \frac{1}{r^3},~
(r>>\ll).
\]
The integral in the right hand side of the equation (\ref{T5}) is
determined by the long- distance region ($S(x)<<1$) and can be
estimated as $O(a^2)$.
Thus the solution of \eq{1} as a function of the parameter
$a$ is found.

On the other hand at the sufficiently small $r$ the function $\phi(x)$
can be found approximately as the solution of the equation
$(-\d^{2}+U(x))\phi(x)=0$ due to the fact that the right hand side of
\eq{T4} is $\sim\rho$. The solution is $\phi_{0}(r)=C\ln(r/\a)$ where
$C$ is an arbitrary constant and $\a$ is the scattering length for the
potential $U(r)$ (the region $r>>\a$ is implied).
In two dimensions the scattering length is defined
by the behavior of the scattering amplitude at low energy which (for
our equation) is given by
\[
{\bar f}(k)=\frac{\pi}{\ln k\a/2 + \gamma - i\pi/2} + O(k\a),
\]
where $|{\bar f}(k)|=(2\pi k)^{1/2}|f(k)|$ is the modified scattering
amplitude, $k$ is the momentum and $\gamma=0.1159$
is the Euler constant.
For the 2D hard-sphere potential the scattering length $\a$ is equal
to the radius of the potential. The correction
$\delta\phi(r)\sim a\rho r^2$ to the solution of the
homogeneous equation can be easily estimated using \eq{T3}. The
correction is $\sim$ 1 at the distances of order of the
correlation length $\ll$ however it is small
($\delta\phi(r)\sim a$) at the distances $r\sim\rho^{-1/2}$.
At $r\sim\rho^{-1/2}$ the equation (\ref{T3}) is still
valid. Comparing the solution given by \eq{T3} with the
function $\phi_{0}(r)$ we get the relation
\be
    C\ln(r/\a) = 1 - a/4\pi + (a/2\pi)\ln \left( \ll r \right)
\label{T6}
\ee
which should be valid in the region $\a<<r<<\ll$ with the logarithmic
accuracy. In particular at $r\sim\rho^{-1/2}$ it is valid with the
accuracy up to the terms of order $\sim a$.
Thus in the leading order in the
small parameter $|\ln\rho\a^2|^{-1}$, $C=a/2\pi$ and the energy is
\be
a = \frac{4\pi}{|\ln\rho\a^2|} +
                    O\left(\frac{1}{|\ln\rho\a^2|^2}\right).
\label{T7}
\ee
\eq{T7} is our final result for the two- dimensional system.

Naively, from the three-particle equation the function $S_3$ can be
estimated by an order of magnitude as
$S(1,2,3)\sim S(13)S(23)+permutations$. Then the contribution of $S_3$
to \eq{1} at the distances $\sim\rho^{-1/2}$
which is determined by the integral
$\rho\int dx_{3}S(13)(\d_{1}\d_{3}-\d_3^2)S(123)$ (see \eq{1.2})
is suppressed due to the
smallness of the function $S(x)$ at these distances.
Actually the three-particle equation can be solved in the leading
order in the expansion parameter (that means that the terms
$\sim S_{2}S_{3}$ in this equation should be neglected)
in the momentum representation
and the corresponding integral in the momentum space can be estimated
(see Appendix).

At the same time at short distances
where the two particles interact strongly the pair function is not
expected to be very different from the solution of the two-body
problem whether or not the function $S_3$ is taken into account.
In the other words although the
function $S_3$ is not necessary small at short distances our estimate
of the right hand side of \eq{T4} as well as the equation (\ref{T6})
are valid by an order of magnitude.
Note that the "momentum" corresponding to the two-particle
equation is of order of the correlation length which characterizes
the screening of the pair wave function due to the other particles.
Hence the result (\ref{T7}) has a simple physical interpretation.
Namely the energy of the two particles interacting with the potential
$U$ located in the two-dimensional volume $V$ is $\sim {\bar f}/V$
where the modified scattering amplitude ${\bar f}$ should be
normalized at $k\sim\rho^{1/2}$. Multiplying the result by the number
of pairs we obtain the result (\ref{T7}). In three dimensions that
corresponds to the first term in \eq{aa}.

The other approach to the description of the dilute bose gas
based on the equations for the
distribution functions related to the function $\Phi$
was proposed in ref.\cite{L}. We will show below that the assumptions
made in ref.\cite{L} can be justified in the framework of our
approach. Let us describe briefly this method.
The $n$-particle distribution function is defined by
\be
g_n(x_1,\ldots x_n)=Z^{-1}V^n\int dx_{n+1}\ldots dx_N
\Phi(x_1,\ldots x_N)
\label{LI1}
\ee
\[
Z=\int dx_{1}\ldots dx_N \Phi(x_1,\ldots x_N).
\]
The energy is related to the pair distribution function
$g_2(x_{12})=g(x)$ by means of
\be
      a=\frac{N-1}{N} \int dx U(x)g(x).
\label{LI2}
\ee
Using \eq{1.0} we obtain the equation for $g(x)$:
\be
\left(-\d^2+U(x)\right)g(x)={1\over 2} E_0 g(x) - \rho \int
dx_3 g_{3}(1,2,3)U(x_{23})
\label{LI3}
\ee
\[
- \frac{1}{2}\frac{(N-2)(N-3)}{V^2}\int dx_3dx_4
g_{4}(1,2,3,4)U(x_{34}).
\]
The first and the third terms in the right- hand side are of order
$\sim N$ while the left- hand side is of order $\sim 1$. In
the leading order the cancellation of these terms take place and the
corrections $\sim 1/N$ are important. The approximation for the
distribution functions $g_3$ and $g_4$ used in ref.\cite{L} to
evaluate the right hand side of \eq{LI3} is the superposition
approximation:
\be
g_n(x_1,\ldots x_n)=\prod_{i<j}^{n}\tilde{g}(x_{ij})
\label{LI4}
\ee
{}From physical considerations it is clear that in the limit
$V\rightarrow\infty$ the function $\tilde{g}(x)\rightarrow g(x)$
while for the finite volume it can be found using the set of the
equations
\[
\int dx_n g_n(x_1,\ldots x_n) = V g_{n-1}(x_1,\ldots x_{n-1}).
\]
For instance, for the four- particle function $g_4$,
substituting the ansatz (\ref{LI4}) into the formula
$\int dx_{3}dx_{4}~g_{4}(1,2,3,4)=V^{2}g(x_{12})$
we find that the corresponding pair function is
\[
\tilde{g}^{(4)}(x)=g(x)\left(1-\frac{1}{V}\int dx_3
f(x_{13})f(x_{23})\right),
\]
where the function $f(x)=1-g(x)$ can be taken in the infinite volume
limit. These corrections are not important in the second term of
the right- hand side of \eq{LI3}.
Substituting the expressions for the functions $g_3$ and
$g_4$ into the equation (\ref{LI3}) and taking the limit
$V\rightarrow\infty$ we get an equation for the pair distribution
function $g(x)$.
Assuming that the function $f(x)$ is small and
retaining the terms of the leading order in $f(x)$ we get
\be
\left(-\d^2+U(x)\right)g(x)=2a\rho f(x)
                      - a\rho^2 \int dx_3 f(x_{13})f(x_{23})
\label{L1}
\ee
This equation is valid for those $r$ where the function $f(x)$ is
small ($f(x)<<1$).

The equation (\ref{L1}) can be solved in the same way as \eq{1}.
For instance the solution of \eq{L1} for 2D problem in the region
where $f(x)<<1$ is
\be
g(x)=1+a/4\pi+(a/2\pi)\ln\left((a\rho)^{1/2}r\right)
\label{L2}
\ee
for $r<<\ll$. Repeating the arguments leading to \eq{T6} we obtain the
result (\ref{T7}).

In 3D the solution of the equation (\ref{L1}) at the distances
$r<<\ll$ is
\[
g(x)=\left(1+\frac{16}{15\pi^2}\a^{3/2}\rho^{1/2}\right)-\frac{a}{r}.
\]
Comparing this function with the solution of the homogeneous equation
$\phi_{0}(r)=C(1-\a/r)$ which is valid with the accuracy
$\sim \a^{3/2}\rho^{1/2}$ at $r<\rho^{-1/3}$
we obtain the result (\ref{aa}).

Let us comment on the relation of our approach to that of
ref.\cite{L}. The function $g(x)$
can be calculated using the obtained wave function of the Jastrow
form. Formally the problem is similar to the calculation of the
distribution functions of the classical liquid.
Although the expansion in density for the corresponding partition
function is not valid (integral $\int dxS(x)$ diverges at large
distances) the formulas of ref.\cite{L} resulting in the \eq{L1} can
be obtained in the lowest order in the expansion parameter.
One can establish the direct correspondence between the equations
(\ref{1}) and (\ref{L1}). The function $g(x)$ at long distances
can be calculated by means of the cluster expansion (for instance, see
\cite{RC}). The connected diagrams with the lines corresponding to the
function $\phi(x)-1$ between the points 1 and 2 are considered.
First, the Fourier transform of the function $\phi(x)-1$
may be replaced by $S_k$. In the integrals corresponding to the
diagrams of the cluster expansion the long distances where the
function $S(x)$ is small are important. Therefore only the diagrams
that does not contain the two different paths connecting the points
1, 2 are required. These diagrams can be refered to as a "chain"
diagrams (see Fig.1). Adding once more line to each of these diagrams
leads to an extra suppression. Since $S(r)\rightarrow 0$
at $r\rightarrow\infty$, it is clear that the sum of the "chain"
diagrams gives the exact result for the asymptotic of the distribution
function. The "chain" diagrams can be summed up in the momentum
representation. We obtain the following relation:
\be
f_k = - \frac{S_k}{1-\rho S_k}.
\label{L3}
\ee
Substituting the solutions of the equations (\ref{1}) and (\ref{L1})
into \eq{L3} we find that this relation is valid with the accuracy
required for the estimates (\ref{aa}) and (\ref{T7}).

The same approximation can be used to calculate the Feynman structure
factor $F(k)$ which determines the energy of the low-lying
excitations (phonons). In this case the distribution function related
to the square of the wave function $\Phi$ (defined as in \eq{LI1})
should be calculated. We find
\[
F(k)=\rho\frac{1}{1-2\rho S_k}.
\]
In contrast to the other wave functions of the Jastrow form used in
the variational studies of the bose liquid \cite{M} we obtain the
correct behavior $F(k)\sim k$ at $k\rightarrow 0$ due to the behavior
$S(r)\sim 1/r^{d-1}$ at the asymptotically large distances \cite{RC}
(for a related discussion see ref.\cite{K}).
The phonon energy
\[
\omega(k)={\rho k^2 \over F(k)} = \sqrt{k^4+4a\rho k^2}
\]
is in agreement with the predictions of the other approaches
\cite{B,LY}.
The momentum distribution $n_k$ can be readily calculated. In terms of
the wave function it has the form
\[
n_k = N\int dx e^{ikx} \int dx_2 \ldots dx_N
\Phi(x,~x_2 \ldots x_N) \Phi(0,~x_2 \ldots x_N)
\]
($\int_{k}n_k=\rho$, the normalization $<\Phi|\Phi>=1$ is implied).
The result is as follows
\[
n_k = \frac{\rho^2 S_k^2}{1-2\rho S_k} =
\frac{1}{2}\left( \frac{k^2+2a\rho}{\sqrt{k^4+4a\rho k^2}} -1\right)
\]
which coincides with the known expression \cite{B,LY}. In two
dimensions the expression (\ref{T7}) should be substituted for $a$. In
2D the condensate is $N_0=N(1-{1 \over 4\pi}a)$.

Finally, the situation is different for one-dimensional problem.
{}From the point of view of our method the low density limit coincide
with the weak coupling limit.
In this case the expression (\ref{Bog}) is correct in the weak
interaction limit regardless of the shape of the potential
while for the strong coupling the perturbation theory is inapplicable
which is in agreement with the exact solution of the problem for the
$\delta$- function potential \cite{LL}.
The same can be true for the lattice system.
For instance for a system described
by the Hamiltonian (for example, see ref.\cite{RK})
\[
H=-t\sum_{<ij>}(b_{i}^{+}b_{j}+h.c.)+U\sum_{i}n_i(n_i-1),
\hspace{0.2in} n_i=b_{i}^{+}b_i,
\]
at $U\rightarrow 0$ and the density of order of unity \eq{Bog} is
asymptotically exact in any dimensions.

In conclusion, we presented the method to describe the interacting
bose gas at zero temperature.
The expansion in the irreducible functions for a logarithm of the
ground state wave function was used.
For a low density the equation for a Jastrow wave function was solved.
It was argued that the contribution of the three-particle component of
the wave function to the equation is suppressed.
For three-dimensional system the leading order correction
for the ground state energy in particle density was reproduced.
For the two-dimensional dilute
bose gas the ground state energy in the leading order in the parameter
$|\ln\a^2\rho|^{-1}$ where $\a$ is two-dimensional scattering length
was obtained.

{\bf Aknowledments}

The author is grateful to D.T.Son for discussions. This work was
supported, in part, by a Soros Foundation Grant awarded by the
American Physical Society.

\vspace{0.4in}
\begin{center}
                  {\bf Appendix}
\end{center}

Here we estimate the contribution of the three-particle function
$S_{3}(x_1,x_2,x_3)$ to the equation (\ref{1}). The three- particle
equation is obtained after the integration of the basic equation
(\ref{1.0}) over the coordinates $4,\ldots N$. It is necessary to take
into account the equations (\ref{1.4}) and (\ref{1}) for the terms
that depend only on one of the variables $x_{12},x_{13},x_{23}$. In
the coordinate space the equation is
\be
\frac{1}{2}(\d_1^2+\d_2^2+\d_3^2)S_3(1,2,3)+ \d_1 S(12)\d_1 S(13)
+ \d_2 S(21)\d_2 S(23)
\label{A1}
\ee
\[
+ \d_1 S(12)\d_1 S(13) +\ldots =\frac{1}{V\rho}
\left(F(12)+F(13)+F(23)\right)
\]
where the dots stands for the terms of order
$\sim S_{2}S_{3},~S_{3}^{2}$. These terms are not important for the
solution of \eq{A1} in the leading order. The right-hand side is equal
to zero in the infinite volume limit. In the momentum representation
the function $S(k_1,k_2,k_3)$ is defined as
\[
\int dx_1 dx_2 dx_3 e^{ik_{1}x_1+ik_{2}x_2+ik_{3}x_3} S(1,2,3) =
 (2\pi)^d \delta^{d}(k_1+k_2+k_3)S(k_1,k_2,k_3).
\]
The solution of the equation (\ref{A1} is
\be
S(k_1,k_2,k_3)=-\frac{2}{k_1^2+k_2^2+k_3^2}\left(
k_{1}k_{2}S_{k_1}S_{k_2}+k_{1}k_{3}S_{k_1}S_{k_3}
+k_{2}k_{3}S_{k_2}S_{k_3} \right).
\label{A2}
\ee
The contribution of the function $S_3$ to the left-hand side of
\eq{1} has the following form:
\be
2\rho\int dx_3 S(13)(\d_1\d_3-\d_3^2)S_3(1,2,3),
\label{A3}
\ee
which is the function of the variable $x_{12}=x$. Substituting the
solution (\ref{A2}) to the integral in the momentum space
corresponding to the Fourier transform of the function (\ref{A3}) we
obtain the following expression:
\be
2\rho \int_{p} \frac{2p^2-kp}{p^2+k^2-kp}S_p \left(
k(k-p)S_{k}S_{k-p}+p(p-k)S_{p}S_{k-p}+kpS_{k}S_p \right),
\label{A4}
\ee
where $k$ is the external momentum. It is possible to estimate the
integral of \eq{A4} at different values of the momentum $k$ with the
help of \eq{T2}. For example at $k=0$ we have
\[
4\rho \int_{p} p^2 S_p^3
\]
This integral is easily estimated as $O(a^2)$ and
$O(a(a^{3/2}\rho^{1/2}))$ respectively
in two- and three- dimensional space. Analysing \eq{A4} it is easy to
show that these estimates are valid at arbitrary $k$. Therefore the
considered contribution does not change result in the approximation
required for the derivation of the formulas (\ref{aa}) and
(\ref{T7}).

\end{document}